\documentclass[12pt]{article}
\usepackage{latexsym}
\usepackage{amssymb}
\usepackage[english]{babel}
\usepackage{amsmath}
\title{The convolution formula for a decay rate}
\author{V.I. Kuksa}
\date{Institute of Physics, Rostov State University,\\
 pr. Stachki 194, Rostov-on-Don, 344090 Russia,\\
 E-mail address: kuksa@list.ru}
\begin{document}

\maketitle
\begin{abstract}

The convolution formula is derived within the framework of the
decay-chain method for decay channels with three and four
particles in a final state. To get this formula exactly for
unstable particles of any type one must modify the propagators of
vector and spinor fields. In this work we suggest proper
modifications and get the convolution formula by direct
calculations. It was noted that this approach naturally arises in
the model of unstable particles with random mass.

PACS number(s): 11.10St, 130000


\end{abstract}

\pagenumbering{arabic} \setcounter{page}{1}

\section{Introduction}

The unstable particles (UP) manifest themselves as resonances
(intermediate states) in the scattering experiments and as
decaying (evolving) states in the decay or oscillation
experiments. The former case is treated with $S$ - matrix or with
renormalized propagator approach. In the second case we must take
into account the instability or finite-width effects (FWE) in a
special way. There are some different methods proposed in
literature to describe the UP as quasi-stable state. The first
method was formulated by Matthews and Salam in Ref. \cite{1},
where an uncertainty principle was taken into consideration. Bohm
et al. \cite{2} suggest the time asymmetric quantum theory of UP
with relativistic Gamov vector, which describes UP in initial or
final states. The model of UP with random (smeared, fuzzy) mass
was suggested in Ref. \cite{3}. This model is based on the
uncertainty relation for the mass of UP and has a close analogy to
the elaboration \cite{1}. An effective theory of UP is discussed
in \cite{4} where the authors have applied the method of
correction factorization.

At last decade the so called convolution method (CM) has been used
to evaluate FWE \cite{5,6}. This method consists in a
semiphenomenological description of invariant mass distribution by
Breit-Wigner-like density function. The convolution formula was
derived within the framework of the model \cite{3} as a direct
consequence of the model approach. This formula was applied in a
phenomenological way to calculate the hadron decay rates \cite{7}.
The FWE (or "mass smearing"\, effects) in hadron decays are very
significant due to large hadron widths. The FWE in the
near-threshold decays $t\rightarrow WZb$, $t\rightarrow cWW,\,
cZZ$ and $A^0(h^0)\rightarrow tbW$ were calculated in Refs.
\cite{5,6} where some analysis of CM applicability is fulfilled.

In this paper we systematically analyze the processes of type
$\Phi \rightarrow \phi_1 \phi \rightarrow \phi_1\phi_2\phi_3$,
where $\phi$ is an UP with a large width and $\Phi, \phi_1,
\phi_2$ are the stable or long-lived particles of any kind. The
prescriptions are suggested for propagators and for polarization
matrices of unstable vector and spinor fields, which exactly lead
to the convolution formula for a decay rate. It was noted that
these prescriptions naturally follow from the model \cite{3}. The
convolution formula for above mentioned decays was got by direct
calculations without any approximations for all type of UP and for
all possible type of $\Phi$ and $\phi_{\alpha}$.

\section{Derivation of the convolution formula for a decay rate}

The convolution formula (CF) can be obtained on the example of
processes $\Phi\rightarrow \phi_1 \phi \rightarrow
\sum_{i,k}\phi_1\phi_i\phi_k$, where $\phi$ is an UP with a large
width. In the general case:
\begin{equation}\label{E:1}
 \Phi\rightarrow\phi_1\phi\rightarrow\phi_1 \sum_{n} X_n\,,
 \end{equation}
where the sum runs over all decay channels of the unstable
particle $\phi$. Some particles of the $X_n=(x_1,x_2,...,x_n)$ can
be unstable too. By means of Eq. (\ref{E:1}) we represent the
connection between the CM and decay-chain method (DCM). When the
sum runs over $(\phi_i, \phi_k)$, which are the stable or
long-lived particles, we have:
\begin{equation}\label{E:2}
  \Gamma(\Phi\rightarrow \phi_1 \phi) = \sum_{i,k}\Gamma(\Phi\rightarrow
  \phi_1 \phi_i \phi_k)\,.
\end{equation}
The CM gives the expression \cite{5}:
\begin{equation}\label{E:3}
 \Gamma(\Phi\rightarrow\phi_1\phi) = \int_{q^2_1}^{q^2_2} \Gamma(\Phi\rightarrow
 \phi_1\phi(q))\rho_{\phi}(q)dq^2\,.
\end{equation}
In Eq. (\ref{E:3}) the value
$\Gamma(\Phi\rightarrow\phi_1\phi(q))$ is the width, calculated in
a stable particle approximation when $m^2_{\phi} = q^2$ and
$\rho_{\phi}(q)$ is some invariant mass distribution function.
From Eq. (\ref{E:1}) it is clear that $\rho_{\phi}(q)$ depends on
all decay-chain of $\phi$, that is, CF can be derived in principle
from the DCM.

When the intermediate state $\phi$ is scalar UP then for any type
of initial $\Phi$ and final $\phi_k$ states a decay width can be
represented in the factored form (see Appendix):
\begin{equation}\label{E:4}
  \Gamma(\Phi\rightarrow\phi_1 \phi_2 \phi_3)=\int_{q^2_1}^{q^2_2}\Gamma(\Phi
  \rightarrow\phi_1 \phi(q))\frac{q\Gamma(\phi(q)\rightarrow\phi_2\phi_3)}
  {\pi \vert P_{\phi}(q)\vert ^2}dq^2\,.
\end{equation}
In Eq. (\ref{E:4}) $q_1=m_2+m_3$ , $q_2=m_{\Phi}-m_1$ and
$P_{\phi}(q)$ is the propagator's denominator of the scalar field
$\phi$. It is noteworthy that the derivation of the convolution
formulae (\ref{E:3}) or (\ref{E:4}) don't depends on the choice of
$P_{\phi}(q)$. This fact is important for a model definition of
$\rho(q)$.

For convenience and completeness we give in Appendix a full list
of the expressions $\Gamma(\Phi\rightarrow\phi_1\phi)$ and
$\Gamma(\Phi\rightarrow\phi_1\phi_2\phi_3)$ for all possible types
of $\Phi$ and $\phi_k$. From Eqs. (\ref{E:2}) and (\ref{E:4}) it
follows that $\rho_{\phi}(q)$ in Eq. (\ref{E:3}) can be expressed
in the form:
\begin{equation}\label{E:5}
 \rho_{\phi}(q)=\frac{q}{\pi\vert P_{\phi}(q)\vert^2}\sum_{i,k}\Gamma(\phi(q)
 \rightarrow\phi_i\phi_k)\equiv\sum_{i,k}\rho^{ik}_{\phi}(q).
\end{equation}

A generalization of Eq. (\ref{E:5}) for the cases when three or
more particles are in a final state is straightforward (see
Appendix). Thus, we can always represent
$\Gamma(\Phi\rightarrow\phi_1\phi)$, where $\phi$ is scalar UP
with a large width, by the convolution formula (\ref{E:3}). If one
uses the parametrization $q\Gamma(q)=Im\Sigma(q)$ and
Dyson-resummed propagator
\begin{equation}\label{E:6}
 P_{\phi}(q)=q^2-m^2_{\phi}(q)-iIm\Sigma_{\phi}(q),\,\,
 m^2_{\phi}(q)=m^2_{0\phi}+Re\Sigma_{\phi}(q),
\end{equation}
then the $\rho_{\phi}(q)$ can be written in the Lorentzian
(Breit-Wigner type) form:
\begin{equation}\label{E:7}
 \rho_{\phi}(q)=\frac{1}{\pi}\,\frac{Im\Sigma_{\phi}(q)}{[q^2-m^2_{\phi}]^2+[Im\Sigma_{
 \phi}(q)]^2}\,.
\end{equation}
The expression (\ref{E:7}) have been used in the previous papers
\cite{5,6,7}.

The situation drastically changes when UP is vector or spinor
field. Traditional propagators
$P_{\mu\nu}(q)=-i(g_{\mu\nu}-q_{\mu}q_{\nu}/m^2_V)/P_V(q)$ for the
vector field $V$ and $\hat{P}(q)=i(\hat{q}+m_{\Psi})/P_{\Psi}(q)$
for the spinor field $\Psi$ do not make it possible to represent
$\Gamma(\Phi \rightarrow\phi_1\phi_2\phi_3)$ in the factored form
(\ref{E:4}). As it was noted in Ref. \cite{6} for the vector field
case the numerator $\eta_{\mu\nu}=-i(g_{\mu\nu}-q_{\mu}
q_{\nu}/m^2_V)$ destroys the factorization and we have an
approximate CF. The same disfactorization takes place for the
fermion propagator $~i(\hat{q}+ m_{\Psi})$ too. In this situation
one should analyze the modification of propagators for vector and
spinor field after Dyson summation. Such an analysis is
complicated due to tensor and operator structure of
$\eta_{\mu\nu}$ and $\hat{q}+m$. There are no unit and strict
definitions of these structures in literature. For example, the
$\eta_{\mu\nu}$ has the structure $g_{\mu\nu}-
q_{\mu}q_{\nu}/(m^2_V-im_V\Gamma^0_V)$ in Ref. \cite{6},
$g_{\mu\nu}-q_{\mu}q_{\nu}/ (m_V-i\Gamma_V/2)^2$ in Ref. \cite{5},
$g_{\mu\nu}-q_{\mu}q_{\nu}/q^2$ in Ref. \cite{8}, and
$(g_{\mu\nu}-q_{\mu}q_{\nu}/q^2)P_T(q^2)+(q_{\mu}q_{\nu}/q^2)P_L(q^2)$
in Ref. \cite{9}. By direct calculation one can check that
$\eta_{\mu\nu}=-i(g_{\mu\nu}-q_{\mu} q_{\nu}/q^2)$ leads to the
factored expression (\ref{E:4}), that is to CF.

The model of UP with random mass \cite{3} contains the general
designation how to modify propagators $P_{\mu\nu}(q)$ and
$\hat{P}(q)$. The "smearing"\, of UP's mass in according to
uncertainty principle leads to the modified dispersion relation
$q^2=\mu^2$, where $\mu$ is nonfixed random mass of UP with some
distribution function $\rho(\mu)$. In another way we have the
"smeared"\, or "fuzzy"\, mass-shell and the general prescription
$m^2\rightarrow \mu^2=q^2$, where $q$ is timelike arbitrary
momentum. Then the modified propagators for vector and spinor
fields are:
\begin{equation}\label{E:8}
 P_{\mu\nu}(q)=-i(g_{\mu\nu}-q_{\mu}q_{\nu}/q^2)/P_V(q),
\end{equation}
and
\begin{equation}\label{E:9}
 \hat{P}(q)=i(\hat{q}+q)/P_{\Psi}(q),\,q=\sqrt{(qq)}\,.
\end{equation}
In Eqs. (\ref{E:8}) and (\ref{E:9}) the $P_V(q)$ and $P_{\Psi}(q)$
are some functions, which don't influence on convolution structure
of the formulae (\ref{E:3}) and (\ref{E:4}). The prescription
(\ref{E:8}) coincides with the definition in Ref. \cite{8}. It
should be noted that the expressions (\ref{E:8}) and (\ref{E:9})
approximately coincide with standard ones at peak vicinity
$q^2\simeq m^2(m)$.

Direct calculations with help of Eqs. (\ref{E:8}) and (\ref{E:9})
(see Appendix) lead to the factored form (\ref{E:4}) for
$\Gamma(\Phi\rightarrow\phi_1\phi_2\phi_3)$ when the UP in an
intermediate state is vector or spinor field. The calculations
were fulfilled for all possible type of particles $\Phi$ and
$\phi_k$. For completeness we give a full list of the expressions
for $\Gamma(\Phi\rightarrow \phi_1\phi_2\phi_3)$ in Appendix.
Thus, the prescriptions (\ref{E:8}) and (\ref{E:9}) always lead to
CF (\ref{E:4}) or (\ref{E:3}) and (\ref{E:5}). The expression
(\ref{E:7}) runs out from Eq. (\ref{E:5}) and parametrization
$q\Gamma(q)=Im\Sigma(q)$, when we use the expression (\ref{E:6}).
It should be noted that applicability of this approach in the
context of gauge theories is limited to low orders because of the
appearance of gauge dependence \cite{8,9,10,11,12}. Moreover, we
should redefine $P_{\Psi}(q)$ for spinor UP with account of
$m_{\Psi}(q)=m_{o\Psi}+Re\Sigma_{\Psi}(q)$ and
$\Gamma_{\Psi}(q)=Im\Sigma_{\Psi}(q)$. However, as was noted
early, the convolution structure is not subject to definition of
$P(q)$.

To get the expression for $\Gamma(\Phi\rightarrow\phi_1\phi(q))$,
where $\phi(q)$ is $V$ or $\Psi$, one need the definitions of
polarization matrixes for vector and spinor fields. With
prescription $m\rightarrow q$ these definitions are:
\begin{equation}\label{E:10}
 \sum_{e} e_{\mu}e^*_{\nu}=-(g_{\mu\nu}-q_{\mu}q_{\nu}/q^2),
\end{equation}
and
\begin{equation}\label{E:11}
 \sum_{\nu} u^{\nu,\pm}_{\alpha}(q)\bar{u}^{\nu,\mp}_{\beta}(q)=
 \frac{(\hat{q}\mp q)_{\alpha\beta}}{2q^0}\,.
\end{equation}
The expressions (\ref{E:8})\,-(\ref{E:11}) are the continuation of
standard ones to the "fuzzy" \,(smeared) mass-shell. The
factorization of $\Gamma(\Phi \rightarrow\phi_1\phi_2\phi_3)$ on
"fuzzy"\, mass-shell results due to the right sides of Eqs.
(\ref{E:10}) and (\ref{E:11}) equal to the numerators of Eqs.
(\ref{E:8}) and (\ref{E:9}). Such a factorization takes place on
the usual mass-shell.

Direct calculations do not lead to the CF in the cases when there
are two or more UP in the intermediate or final states. In the
framework of the model \cite{3} for $\Phi\rightarrow\phi_1\phi_2$,
where $\phi_1$ and $\phi_2$ are the UP with large width, we get
the expression for the width, which is usually applied in a
phenomenological way:
\begin{equation}\label{E:12}
 \Gamma(\Phi\rightarrow\phi_1\phi_2)=\int dq^2_1\rho_1(q_1)\int dq^2_2\rho_2(q_2)
 \Gamma(\Phi\rightarrow\phi_1(q_1)\phi_2(q_2)).
\end{equation}
Representation of the width
$\Gamma(\Phi\rightarrow\phi_1\phi_2\phi_3)$ in the form
(\ref{E:4}) is actually the transition from DCM to CM. Such
transitions are very complicated in the cases when there are many
unstable and stable particles and there are many sections in the
decay-chain. Instead of transition from DCM to CM we can derive
the convolution formula in the framework of the model \cite{3} for
general case. In Ref. \cite{3} such a derivation was done in the
case of scalar UP, therefore the discussion of this problem for
the vector and spinor UP is actual.

\section{Summary}

In this paper we have demonstrated in detail the connection
between DCM and CM for the decay channels with three-particle
final states. The convolution formula was derived with help of the
prescription $m\rightarrow q$ by direct calculations for all types
of UP. It was noted that this prescription naturally arises in the
model of UP with random mass \cite{3}. Transition from DCM to CM
is very complicated in the case of many-particle decay-chain.
Therefore an alternative proof of CF in the general case is
actual.

\section{Appendix}

In this section we use the $\tilde{\lambda}$-function, which
describes the kinematics of the process
$\phi(q)\rightarrow\phi_1(k_1)\phi_2(k_2)$. In the $\vec{q}=0$
frame of reference:
\begin{equation}\label{E:13}
 \tilde{k}_{\alpha}=\frac{1}{2}q\tilde{\lambda}(k_1,k_2;q),\,\tilde{k}_{\alpha}
 \equiv \vert\vec{k}_{\alpha}\vert,\,\alpha=1,2\,,
\end{equation}
where the $\tilde{\lambda}(k_1,k_2;q)$ is defined in analogy with the known Kallen
function $\lambda(k_1,k_2;q)=\\q^2\tilde{\lambda}^2(k_1,k_2;q)$:
\begin{equation}\label{E:14}
 \tilde{\lambda}(k_1,k_2;q)=[1-2\,\frac{k^2_1+k^2_2}{q^2}+\frac{(k^2_1-k^2_2)^2}
 {q^4}]^{1/2}\,.
\end{equation}
The expressions $\Gamma_i(\phi\rightarrow\phi_1\phi_2)$ for all
types of particles can be represented in the form:
\begin{equation}\label{E:15}
 \Gamma_i(\phi\rightarrow\phi_1\phi_2)=\frac{g^2}{8\pi}\tilde\lambda(m_1,m_2;m)
 f_i(m_1,m_2;m),
\end{equation}
where $f_i(m_1,m_2;m)$ depends on the interaction Lagrangian. To illustrate CM
we use the simplest Lagrangians:
\begin{equation}\label{E:16}
 L_k=g\phi\phi_1\phi_2;\,g\phi V_{\mu}V^{\mu};\,g\phi\bar{\psi}_1\psi_2;\,
 gV_{\mu}(\phi^{,\mu}\phi_1-\phi_1^{,\mu}\phi);\,gV_{\mu}\bar{\psi}_1\gamma^{\mu}
 (c_V+c_A\gamma_5)\psi_2;\,g\phi V^{\mu}_1 V_{2\mu}.
\end{equation}
In Eq. (\ref{E:16}) $\phi ,V_{\mu}$ and $\psi$ are the scalar,
vector and spinor fields. Then the $f_i(m_1,m_2;m)$ in Eq.
(\ref{E:15}) are defined by the following expressions:
\begin{align}
 &\phi\rightarrow\phi_1\phi_2,\,\,\,f_1=\frac{1}{2m}.\\
 &\phi\rightarrow V_1 V_2,\,\,\,     f_2=\frac{1}{m}[1+\frac{(m^2-m_1^2-m_2^2)^2}
                                       {8m_1^2 m_2^2}].\\
 &\phi\rightarrow\psi_1\psi_2,\,\,\, f_3=m[1-\frac{(m_1+m_2)^2}{m^2}].\\
 &\phi\rightarrow\phi_1 V,\,\,\,     f_4=\frac{1}{2}m\frac{m^2}{m_2^2}                                       \tilde{\lambda}^2 (m_1,m_2;m),\,\,m_2=m_V.\\
 &V\rightarrow\phi_1\phi_2,\,\,\,    f_5=\frac{1}{6}m\tilde{\lambda}^2                                        (m_1,m_2;m),\,\,m=m_V.\\
 &V\rightarrow\psi_1\psi_2,\,\,\,    f_6=\frac{2}{3}m\{(c^2_V+c^2_A)[1-\frac{m^2_1+m^2_2}                                        {2m^2}-\frac{(m^2_1+m^2_2)^2}{2m^4}]+                                        3(c^2_V-c^2_A)\frac{m_1m_2}{m^2}\}.\\
 &V\rightarrow V_1\phi,\,\,\,        f_7=\frac{1}{3m}[1+\frac{(m^2+m^2_1-m^2_2)^2}                                        {8m^2m^2_1}],\,\,m_2=m_{\phi}.\\
 &V\rightarrow V_1 V_2,\,\,\,        f_8=\frac{1}{24}\frac{m^5}{m^2_1m^2_2}\{1+                    8(\mu_1+\mu_2)-2(9\mu_1^2+16\mu_1\mu_2+9\mu_2^2)
 +8(\mu_1^3-4\mu_1^2\mu_2\\&-4\mu_1\mu_2^2+\mu_2^3)                   +\mu_1^4+8\mu_1^2\mu_2^2+8\mu_1\mu_2^3+\mu_2^4\},\,\,                                        \mu_{\alpha}=\frac{m^2_{\alpha}}{m^2}.\notag\\
 &\psi\rightarrow\phi\psi_1,\,\,\,   f_9=\frac{1}{2}m(1+2\frac{m_1}{m}+\frac                                        {m^2_1-m^2_2}{m^2}),\,\,m=m_{\psi},m_2=m_{\phi}.\\
 &\psi\rightarrow\psi_1 V,\,\,\,     f_{10}=(c^2_V+c^2_A)\frac{(m^2-m^2_1)^2+                                         m^2_2(m^2+m^2_1-2m^2_2)}{2m m^2_2}-                                         3m_1(c^2_V-c^2_A),\\&m=m_{\psi},m_2=m_V.\notag
\end{align}

Using the expressions (17)\,-(26) we can represent
$\Gamma(\Phi\rightarrow\phi_1\phi_2\phi_3)$ in a compact and
universal form for all types of decay channels. Here we shortly
describe the method of $\Gamma(\Phi\rightarrow\phi_1\phi_2\phi_3)$
calculation. This value can be always written as:
\begin{equation}\label{E:27}
 \Gamma=\frac{k}{p^0}\int J(\vert M(k_i,m_i)\vert^2)\frac{d\bar{k}_1}{k^0_1}\,,
\end{equation}
where $M(k_i,m_i)$ is an amplitude, $p$ and $k_i$ are momentum of
$\Phi$ and $\phi_i$, $k$ is some numerical factor, and
\begin{equation}\label{E:28}
 J(\vert M\vert^2)=\int \vert M\vert^2\delta(p-k_1-k_2-k_3)\frac{d\bar{k}_2d\bar{k}_3}
 {k^0_2k^0_3}.
\end{equation}
The integral $J(\vert M\vert^2)$ is calculated in $\bar{q}=0$ frame of reference
and as a result we have the noncovariant expression
\begin{equation}\label{E:29}
 J(\vert M\vert^2)\,\longrightarrow\,f(q^0,q^0 p^0,q^0 k^0_3,\bar{p}^2,...).
\end{equation}
This expression can be always reconstructed to covariant form using $\bar{q}=0$:
\begin{equation}\label{E:30}
 q^0\rightarrow q=\sqrt{(qq)},\,q^0p^0\rightarrow (qp),\,q^0k^0_1\rightarrow (qk_1),\,
 \bar{p}^2=(p^0)^2-m^2\rightarrow (pq)^2/q^2-m^2,...
\end{equation}
Then we pass to the $\bar{p}=0$ frame of reference and change the
variable in Eq. (\ref{E:27}) according to
\begin{equation}\label{E:31}
 \frac{d\bar{k}_1}{k^0_1}=-\frac{1}{2m}\tilde{k}_1 dq^2 d\Omega=-\frac{1}{4}
 \tilde{\lambda}(q,m_1;m)dq^2d\Omega.
\end{equation}

Using this simple method and prescriptions (\ref{E:8}), (\ref{E:9}) we have got
by tedious but straightforward calculations the general expression for
$\Gamma(\Phi\rightarrow\phi_1\phi_2\phi_3)$, when $\Phi,\phi$ and $\phi_k$ are
all of possible type particles:
\begin{equation}\label{E:32}
 \Gamma_{\alpha\beta}(\Phi\rightarrow\phi_1\phi_i\phi_k)=\frac{g^2_1g^2_2}{2^6\pi^3}
 \int_{q^2_1}^{q^2_2}\tilde{\lambda}(q,m_1;m)f_{\alpha}(q,m_1;m)\tilde{\lambda}
 (m_i,m_k;q)f_{\beta}(m_i,m_k;q)\frac{qdq^2}
 {\vert P_{\phi}(q)\vert^2}\,,
\end{equation}
where $q_1=m_i+m_k$ and $q_2=m-m_1$. From Eqs. (\ref{E:32}) and
(\ref{E:15}) it follows:
\begin{equation}\label{E:33}
 \Gamma_{\alpha\beta}(\Phi\rightarrow \phi_1\phi_i\phi_k)=\int_{q^2_1}^{q^2_2}dq^2
 \Gamma_{\alpha}(\Phi\rightarrow\phi_1\phi(q))\frac{q\Gamma_{\beta}(\phi(q)
 \rightarrow\phi_i \phi_k)}{\pi\vert P_{\phi}(q)\vert^2}\,.
\end{equation}
In the approximation
\begin{equation}\label{E:34}
 \Gamma(\Phi\rightarrow\phi_1\phi)=\sum_{i,k}\Gamma(\Phi\rightarrow\phi_1\phi_i
 \phi_k)
\end{equation}
we get the known convolution formula
\begin{equation}\label{E:35}
 \Gamma(\Phi\rightarrow\phi_1\phi)=\int_{q^2_1}^{q^2_2}\Gamma(\Phi\rightarrow
 \phi_1\phi(q))\rho_{\phi}(q)dq^2\,,
\end{equation}
where
\begin{equation}\label{E:36}
 \rho_{\phi}(q)=\frac{q}{\pi\vert P_{\phi}(q)\vert^2}\sum_{i,k}\Gamma(\phi(q)\rightarrow \phi_i\phi_k).
\end{equation}
The same result can be received for many-particle decay channels of UP $\phi\rightarrow
\phi_1\phi_2\phi_3...$\,. For example, let us consider the decay chain $\Phi
\rightarrow\phi_1\phi\rightarrow\phi_1\phi_2\phi_3\phi_4$, where $\phi_k$ are the
scalar fields. Then for the simplest contact interaction we have:
\begin{equation}\label{E:37}
 \Gamma_{\Phi}=\frac{g^2_1g^2_2}{2^{13}\pi^8 p^0}\int\frac{d\bar{k}_1}
 {k^0_1\vert P_{\phi}(q)\vert^2}\int\int\int\delta(q-k_2-k_3-k_4)\frac
 {d\bar{k}_2d\bar{k}_3d\bar{k}_4}{k^0_2k^0_3k^0_4}\,,
\end{equation}
where $q=p-k_1$ and
\begin{equation}\label{E:38}
 \Gamma_{\phi}(q)\equiv \Gamma(\phi(q)\rightarrow\phi_1\phi_2\phi_3)=
 \frac{g^2_2}{2^9\pi^5q^0}\int\int\int\delta(q-k_2-k_3-k_4)\frac
 {d\bar{k}_2d\bar{k}_3d\bar{k}_4}{k^0_2k^0_3k^0_4}\,.
\end{equation}
From Eqs. (\ref{E:37}), (\ref{E:38}) and (\ref{E:17}) it follows:
\begin{equation}\label{E:39}
 \Gamma_{\Phi}=\int_{q^2_1}^{q^2_2}dq^2\Gamma_{\Phi}(q)\frac{q\Gamma_{\phi}(q)}
 {\pi\vert P_{\phi}(q)\vert^2}\,,
\end{equation}
where
$\Gamma_{\Phi}(q)\equiv\Gamma(\Phi\rightarrow\phi_1\phi(q))$.
Using the factorizable $\vert M\vert^2$ we can get the result
(\ref{E:39}) by direct calculations for others types of particles
$\phi_k$. It should be noted that the factored (\ref{E:33}) and
convolution (\ref{E:35}) structures take place for any choice of
$P_{\phi}(q)$.


\begin{thebibliography}{9}
 \bibitem{1}
P.T. Matthews and A. Salam, Phys. Rev. 112 (1958) 283;\\
P.T. Matthews and A. Salam, Phys. Rev. 115 (1959) 1079.
 \bibitem{2}
A. Bohm, N.L. Harshman, H. Kaldass, S. Wickramasekara, Eur. Phys.
J. C 18 (2000) 333;\\
A.R. Bohm and Y. Sato, hep-ph/0412106;\\
A.R. Bohm, N.L. Harshman, Nucl. Phys. B 581 (2000) 91.
 \bibitem{3}
V.I. Kuksa, Semyphenomenological model of unstable particles, in
Proceeding of the 17 International Workshop, Samara-Saratov,
Russia, September 4-11, 2003, edited by M. Dubinin, V. Savrin
(D.V. Skobeltsyn Institute of Nuclear Physics, Moskow State
University), p.350;\\
V.I. Kuksa, hep-ph/0404281.
 \bibitem{4}
M. Beneke, A.P. Chapovsky, A. Signer and G. Zanderighi,hep-ph/0312331;\\
G.Zanderighi, hep-ph/0405124.
 \bibitem{5}
G. Mahlon, S. Parke, Phys. Lett. B 347 (1995) 394;\\
G. Mahlon, S. Parke, hep-ph/941225;\\
G. Mahlon, hep-ph/9810485;\\
G. Altarelli, L. Conti, V. Lubicz, Phys. Lett. B 502 (2001) 125.
 \bibitem{6}
S. Bar-Shalom, G. Eilam, M. Frank and I. Turan, hep-ph/0506167.
 \bibitem{7}
A.N. Kamal and R.C. Verma, Phys. Rev. D 45 (1992) 982;\\
T. Uppal and R.C. Verma, Z. Phys. C 56 (1992) 273;\\
T. Uppal and R.C. Verma, Phys. Rev. D 46 (1992) 2982;\\
H. Kaur and M.P. Khanna, J. Phys. G: Nucl. Part. Phys. 26 (2000)
387.
 \bibitem{8}
B.A. Kniehl, A. Sirlin, Phys. Lett. B 530 (2002) 129.
 \bibitem{9}
R.G. Stuart, Phys. Rev. Lett. 70 (1993) 3193.
 \bibitem{10}
B.A. Kniehl and A. Sirlin, Phys. Rev. Lett. 81 (1998) 1373;\\
B.A. Kniehl and A. Sirlin, Phys. Lett. B 440 (1998) 136.
 \bibitem{11}
P.A. Grassi, B.A. Kniehl and A. Sirlin, Phys. Rev. Lett. 86 (2001)
389;\\
P.A. Grassi, B.A. Kniehl and A. Sirlin, Phys. Rev. D 65 (2002)
085001.
 \bibitem{12}
 M.L. Nekrasov, Phys. Lett. B 531 (2002) 225.
\end{thebibliography}
\end{document}